\documentclass[twocolumn]{aa}
\usepackage{graphicx}
\usepackage{txfonts}
\newcommand{\hii}{H\,{\sc ii}}

\newcommand{\heii}{He\,{\sc ii}}

\newcommand{\feii}{Fe\,{\sc ii}}
\newcommand{\feiii}{Fe\,{\sc iii}}

\newcommand{\cii}{C\,{\sc ii}}
\newcommand{\ciii}{\ifmmode {\rm C}\,{\sc iii} \else C\,{\sc iii}\fi}
\newcommand{\civ}{\ifmmode {\rm C}\,{\sc iv} \else C\,{\sc iv}\fi}

\newcommand{\niii}{N\,{\sc iii}}
\newcommand{\niv}{N\,{\sc iv}}
\newcommand{\nv}{N\,{\sc v}}
\newcommand{\oi}{O\,{\sc i}}

\newcommand{\oiii}{O\,{\sc iii}}

\newcommand{\oiv}{O\,{\sc iv}}

\newcommand{\ovi}{O\,{\sc vi}}

\newcommand{\siiii}{Si\,{\sc iii}}
\newcommand{\Sizw}{Si\,{\sc ii}}
\newcommand{\siiv}{Si\,{\sc iv}}

\newcommand{\aliii}{Al\,{\sc iii}}

\begin{document}
   \title{Elemental Abundances in the Broad Emission Line Region of Quasars 
          at Redshifts larger than 4
         \thanks{Based on observations collected at the VLT-UT1 operated on 
                 Cerro Paranal (Chile) by the European Southern Observatory}
}

   \author{M. Dietrich\inst{1,2,3},
          I. Appenzeller\inst{3},
          F. Hamann\inst{1},
          J. Heidt\inst{3},
          K. J\"{a}ger\inst{4},
          M. Vestergaard\inst{5},
          \and
          S.J. Wagner\inst{3},
          }

   \offprints{M. Dietrich, dietrich@chara.gsu.edu} 

   \institute{Department of Astronomy, University of Florida, 211 Bryant Space
              Science Center, Gainesville, FL 32611-2055, USA\\
              \email{dietrich@astro.ufl.edu, 
                     hamann@astro.ufl.edu}
         \and{current address: Department of Physics \& Astronomy, Georgia
              State University, One Park Place South SE, Suite 700, Atlanta, 
              GA 30303, USA}
         \and
              Landessternwarte Heidelberg--K\"{o}nigstuhl, K\"{o}nigstuhl 12, 
              D--69117 Heidelberg, Germany\\
 \email{iappenze@lsw.uni-heidelberg.de,jheidt@lsw.uni-heidelberg.de,swagner@lsw.uni-heidelberg.de}
         \and 
              Universit\"{a}tssternwarte G\"{o}ttingen, Geismarlandstra\ss e 
              11, D--37083 G\"{o}ttingen, Germany\\
              \email{jaeger@uni-sw.gwdg.de}
         \and
              Department of Astronomy, The Ohio State University, 140 West 
              18th Av., Columbus, OH 43210-1173, USA\\
              \email{vester@astronomy.ohio-state.edu}
             }

   \date{Received September 16, 2002; accepted November 19, 2002}

   \abstract{
We present observations of 11 high redshift quasars ($3.9 \la z \la  5.0$) 
observed with low spectral resolution in the restframe ultraviolet using 
FORS\,1 at the VLT UT\,1.
The emission-line fluxes of strong permitted 
and intercombination ultraviolet emission lines
are measured to estimate the chemical composition of the line emitting gas. 
Comparisons to photoionization calculations indicate gas metallicities in
the broad emission line region in the range of solar to several times solar.
The average of the mean metallicity of each high-z quasar in this sample is
 $Z/Z_\odot = 4.3 \pm 0.3$.
Assuming a chemical evolution time scale of 
$\tau _{evol} \simeq 0.5 - 0.8$\,Gyrs, we derive a redshift of 
$z_f \simeq 6 ~{\rm to}~ 8$ for the onset of the first major star formation 
episode (H$_o = 65$ km\,s$^{-1}$\,Mpc$^{-1}$, $\Omega _M = 0.3$, 
$\Omega _\Lambda = 0.7$), corresponding to an age of the universe of
several $10^8$\,yrs at this epoch.
We note that this epoch is also supposed to be the era of re-ionization 
of the universe.
   }

 \titlerunning{High redshift quasars}
 \authorrunning{M.Dietrich et al.}
   \maketitle

\section{Introduction}
Quasars are among the most luminous objects known in the universe. 
Due to their high luminosity they are excellent tools to probe their galactic 
environment up to the highest redshifts.
There is growing evidence that quasar activity and the formation of their host
galaxies, in particular of massive spheroidal systems, are closely related.
The detection of dark massive objects (DMOs) in the center of nearly every 
galaxy with a significant spheroidal subsystem provides further strong support
for the relationship between galaxy formation and quasar activity (Kormendy \&
Richstone 1995; Magorrian et al.\,1998; Kormendy \& Gebhardt 2001). 
It has been shown that the mass of the DMOs, most likely supermassive black 
holes, is closely correlated to the bulge mass of the host 
galaxy (Gebhardt et al.\,2000; Kaspi et al.\,2000; Merritt \& Ferrarese 2001; 
Tremaine et al.\,2002).
The evolution of the star formation rate indicates that it was more than one 
order of magnitude larger at epochs $z\ga 1$ than in the local universe
(Lilly et al.\,1996; Connolly et al.\,1997; Tresse \& Maddox 1998; 
 Steidel et al.\,1999). 
Strong evidence for the relationship of quasar activity to galaxy formation
accompanied by intense star formation is provided by the detection of large 
amounts of dust ($\sim 10^8$\,M$_\odot$) and molecular gas 
($\sim 10^{10}$\,M$_\odot$) in high redshift quasars
(Andreani, La\,Franca, \& Cristiani 1993; Isaak et al.\,1994;
 Omont et al.\,1996,\,2001; Carilli et al.\,2001).
Several galaxy evolutionary models have been suggested which show that 
galactic spheroids can easily reach solar or supersolar gas-phase 
metallicities on time scales shorter 
than $\sim 1$\,Gyr (Arimoto \& Yoshii 1987; Hamann \& Ferland 1993;
Gnedin \& Ostriker 1997; Fria\c{c}a \& Terlevich 1998; Cen \& Ostriker 1999; 
Salucci et al.\,1999; Kauffmann \& Haehnelt 2000; Granato et al.\,2001; 
Romano et al.\,2002).
As a result, quasars at high redshift are valuable probes for dating the
first star formation in the early universe.
In particular, $z\ga 4$ quasars probe a cosmic era when the universe had an 
age of less than $\sim 10$\,\%\ of its current age (assuming 
$H_o = 65$ km\,s$^{-1}$ Mpc$^{-1}$, $\Omega _M = 0.3$, 
$\Omega _\Lambda = 0.7$).

The prominent emission-line spectrum of quasars contains valuable information 
to estimate the gas metallicity at early epochs due to star formation (for a 
review, see Hamann \& Ferland 1999).
Early studies to estimate the abundances in broad emission line region (BELR) 
gas were based on several generally weak intercombination lines like 
\niv ]$\lambda 1486$, \oiii ]$\lambda 1663$, \niii ]$\lambda 1750$, and
\ciii ]$\lambda 1909$ (Shields 1976; Davidson 1977; Baldwin \& Netzer 
1978; Osmer 1980; Gaskell, Shields, \& Wampler 1981; Uomoto 1984)
and indicated already higher than solar metallicity for the BELR gas. 
Recent studies of the emission line and intrinsic absorption line properties
of ($z\simeq 3$) quasars provide evidence for enhanced metallicities up to an 
order of magnitude above solar (Hamann \& Ferland 1992, 1993; 
Petitjean et al.\,1994; Ferland et al.\,1996; Hamann 1997; Pettini 1999; 
Dietrich et al.\,1999; Dietrich \& Wilhelm-Erkens 2000; Hamann et al.\,2002; 
Warner et al.\,2002; Dietrich et al.\,2002 in prep.).
These high metallicities require a preceding intense star formation phase.

To estimate the chemical composition of the gas in quasar BELRs, nitrogen as a
secondary element is of particular interest. Providing that the secondary 
nitrogen production, i.e., the synthesis of nitrogen 
from existing carbon and oxygen via CNO burning in stars (Tinsley 1979,\,1980;
Wheeler, Sneden, \& Truran 1989), is the dominant source for nitrogen, we can
expect the relation $N/O \sim O/H \sim Z$, i.e., $N/H \sim Z^2$. This scaling 
of $N/H$ with metallicity has been confirmed for many \hii -regions 
(Shields 1976; Pagel \& Edmunds 1981; van Zee et al.\,1998; 
Izotov \& Thuan 1999).

As suggested by Shields (1976), and later developed by Hamann \& Ferland 
(1992,\,1993) and Ferland et al.\,(1996), emission line ratios involving 
\nv $\lambda 1240$ are especially valuable.
Generally, \nv $\lambda 1240$ is stronger than expected in the spectra of high
redshift quasars compared to predictions of standard photoionization models
assuming solar metallicity. 
Hamann \& Ferland (1992,\,1993) showed that 
\nv $\lambda 1240$/\civ $\lambda 1549$ and 
\nv $\lambda 1240$/\heii$\lambda 1640$ 
are useful metallicity indicators. 
Recently, Hamann et al.\,(2002) \mbox{presented} results of a detailed study 
of the 
effects of metallicity and the spectral shape of the photoionizing continuum 
on emission line ratios. They revised the metallicity dependence of line 
ratios involving intercombination lines, recombination lines, and 
collisionally excited lines. They suggest that the most robust indicators of 
the gas chemical composition are the line ratios  
\niii ]$\lambda 1750$/\oiii ]$\lambda 1663$ and 
\nv $\lambda 1240$/(\ovi $\lambda 1034 +$\civ $\lambda 1549$).

We observed a small sample of high redshift quasars ($z\ga 4$) with the 
{\it Very Large Telescope (VLT)} to extend 
prior studies to higher redshift and hence to earlier epochs in the cosmic 
evolution, approaching an age of the universe of $\sim 1$\,Gyr. 
This sample is supplemented with the observation of SDSS\,0338+0021
($z=5.0$; Dietrich et al.\,1999; Fan et al.\,1999).

In section 2 we describe the observations and the data analysis.
In section 3 we present the results of the analysis of the emission line
spectra. We estimate the elemental abundance of the line emitting gas based on 
the line ratios of several diagnostical emission lines (Hamann et al.\,2002).
The overall mean metallicity of the high redshift quasars amounts to
$Z/Z_\odot = 4.3 \pm 0.3$.
These results are discussed and are compared with recent studies 
in section 4. 
The chemical composition of the BELR gas provides further evidence that the 
first major star formation epoch started at a redshift of 
$z_f \simeq  6 ~{\rm to}~ 8$, corresponding to an age of the universe of 
several $10^8$\,yrs. This result is in 

   \begin{table}
      \caption[]{The high-z quasar sample}
         \label{obslog}
     $$ 
         \begin{array}{p{1.0\linewidth}l}
   \begin{tabular}{lcc}
            \hline
            \noalign{\smallskip}
quasar&$m_r$&z\\
            \noalign{\smallskip}
            \hline
            \noalign{\smallskip}
Q 0046-293    &19.38&4.01\\
Q 0101-304    &20.06&4.07\\
SDSS 0338+0021&21.7 &5.00\\
PKS 1251-407  &19.90&4.46\\
APM 1335-0417 &19.40&4.38\\
BRI 1500+0824 &18.84&3.95\\
BRI 1557+0313 &19.80&3.89\\
Q 2133-4311   &20.85&4.18\\
Q 2133-4625   &20.96&4.18\\
Q 2134-4521   &20.15&4.37\\
PC 2331+0216  &19.98&4.09\\
            \noalign{\smallskip}
            \hline
\end{tabular}
         \end{array}
     $$ 
   \end{table}

\noindent
good agreement with recent model predictions relating quasar activity with the
formation of early type galaxies. 
We assumed $H_o = 65$\,km\,s$^{-1}$\,Mpc$^{-1}$, $\Omega _M = 0.3$, and
$\Omega _\Lambda = 0.7$.

\section{Observations and Data Analysis}

We observed the restframe wavelength region $\sim 800 - 2000$\,\AA\ for 11 
quasars with $z\ga 4$ (Table 1), which contains the valuable diagnostic 
ultraviolet emission lines (e.g., \ovi $\lambda 1034$, Ly$\alpha $, 
\nv $\lambda 1240$,  \niv ]$\lambda 1486$, \civ $\lambda 1549$, 
\heii $\lambda 1640$, \oiii ]$\lambda 1663$, \niii ]$\lambda 1750$, and 
\ciii ]$\lambda 1909$).
The quasars were observed on Dec. 26, 1998, July 16 -- 20, and August
14 -- 15, 1999. The total exposure times are ranging from 15 to 60 
minutes, typically 30 minutes.
All observations were carried out using FORS\,1 (focal reducer and 
low-dispersion spectrograph; M\"{o}hler et al.\,1995) at the Cassegrain 
focus of the VLT UT\,1 Antu.
The observations were performed under non-photometric conditions.
The seeing was $\sim 1$\,\arcsec\ on December, and varied between 
0\farcs 8 and 2\farcs 6 during the observations in July and between 
1\farcs 0 and 1\farcs 9 in August.
A Tektronix CCD detector with $2048 \times 2048$ pixels (pixel size 
$24\mu m \times 24 \mu m$) and grating G150I (230\,\AA /mm) were used in 
connection with a blocking filter GG435 to suppress contamination of the 
spectra by the second order for $\lambda \ga 8700$\,\AA . 
With this setting an observed wavelength range of approximately 
$4200 - 10500$\,\AA\ was achieved. 
The quasar spectra were recorded in the multi-object spectroscopy (MOS) mode, 
in the $10^{th}$ ($1\arcsec \times 20 \arcsec$) or 11$^{th}$ 
($1\arcsec \times 22\arcsec$) of the 19 slitlets. 
The position angle of the slit was set perpendicular to the horizon to 
minimize light losses caused by differential refraction. With the exception 
of APM\,1335-0417, the quasars were observed close to the meridian 
(airmass less than $\sim 1.1$).
The other slits were used to observe spectra of faint objects in the close 
environment of the quasars which appear slightly extended on the 
{\it Palomar Observatory Sky Survey} (POSS). 
These serendipity data will be presented and discussed in an upcoming paper.
The standard stars EG\,21, LTT\,1788, and LTT\,7379 (Hamuy et al.\,1992) were 
observed for relative flux calibration each night, with exposure times of
typically 60 seconds.

   \begin{figure*}
   \centering
   \includegraphics[width=145mm]{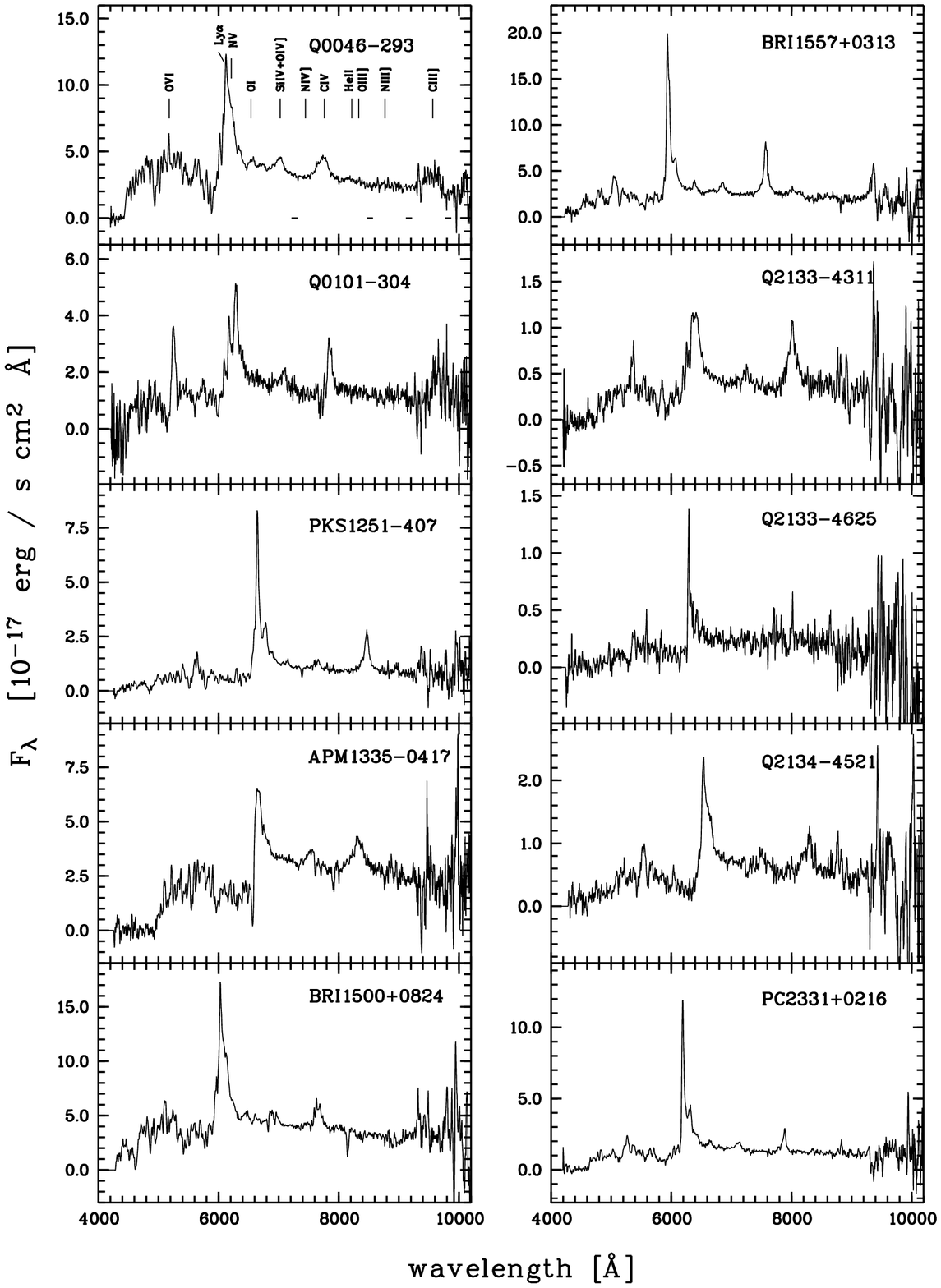}
   \caption{The spectra of the quasars in the observers frame. The flux is 
            given in units of 
            $10^{-17}$ erg s$^{-1}$ cm$^{-2}$ \AA $^{-1}$. Emission lines used 
            in this study are labeled in the spectrum of Q\,0046-293. The 
            horizontal bars indicate the location of the continuum windows used
            to fit the continuum.}
            \label{Fig1}
    \end{figure*}

The quasar and standard star spectra were processed using standard 
MIDAS\footnote{Munich Image Data Analysis System, trade-mark of the European
Southern Observatory} software. 
Cosmic-ray events were removed manually by comparing multiple exposures for
each object.
The night sky component of the 2\,D-spectra was subtracted by fitting third 
order Legendre polynomials perpendicular to the dispersion, along each spatial
row of the spectra using areas on both sides of the object spectrum which were
not contaminated by the quasar or other objects.
The 1\,D-spectra were extracted using an optimal extraction algorithm 
(Horne 1986). 
The width of the spatial extraction windows was adjusted to match the seeing 
recorded during the observation.
The helium-argon wavelength calibration frames, taken for each quasar
MOS-setting, yield a dispersion of 5.4\,\AA/pxl with an internal error of 
$\Delta \lambda \simeq 0.4$\,\AA .
The strong night sky emission lines [\oi ] $\lambda 5577$, 
[\oi ] $\lambda 6300$,  and [\oi] $\lambda 6364$ indicate an absolute 
uncertainty of $\Delta \lambda \simeq 1.1$\,\AA\ 
($\Delta {\rm v} \simeq 50$\,km\,s$^{-1}$).
The FWHM spectral resolution measured from these lines is
$\Delta \lambda \simeq 25$\,\AA .

We corrected each quasar spectrum for the atmospheric c-band, b-band, and
A-band absorption, caused by $O_2$, and the atmospheric water vapor 
absorption bands ($\lambda \lambda 7140 - 7340$\,\AA , $8140 - 8350$\,\AA , 
$9250 - 9600$\,\AA ), using the standard stars observed during the same nights.
The spectra were corrected for atmospheric extinction applying the standard 
curve of La Silla (Schwarz \& Melnick 1993) and for 
interstellar extinction using the $E_{B-V}$ values of Burstein \& Heiles (1982)
and the extinction curve of Savage \& Mathis (1979).

The sensitivity functions, based on individual standard stars, differ by less
than $\sim 4$\,\%\ from the mean sensitivity function for all nights.

\section{Results}

The flux calibrated quasar spectra are shown in Figure 1, with the observed 
flux in units of $10^{-17}$ erg\,s$^{-1}$\,cm$^{-2}$\,\AA$^{-1}$.
The strongest lines in the spectra are Ly$\alpha \lambda 1216$ and 
\civ $\lambda 1549$.
The broad and moderately strong $\lambda 1400$\,\AA\ feature which consists of 
the \siiv $\lambda 1397,1403$ and the \oiv ]$\lambda 1402$ multiplet is quite 
prominent in these quasar spectra.
This feature will be refered as \siiv $\lambda 1400$ in the following.
In addition to these emission lines, several important diagnostic lines like 
\ovi $\lambda 1034$, \heii $\lambda 1640$, \oiii ]$\lambda 1663$, 
\niii ]$\lambda 1750$, and \ciii ]$\lambda 1909$ are visible and marked in
Figure 1.

The observed quasar spectra were transformed to their restframe using the 
redshifts given in Table 1.
To determine the redshift we fit a Gaussian profile to the upper part of the 
the \civ $\lambda 1549$ emission line ($I_\lambda \geq 50$\,\%\ of the peak 
intensity). 

We employed a multicomponent fit to the quasar spectra to determine the 
power-law continuum, $F_{\nu } \propto \nu ^{\alpha}$, the contribution of 
\feii\ and \feiii\ emission, and the weak contribution of the Balmer continuum 
emission (see Dietrich et al.\,2002a,b). 
The power-law continuum was fitted using small spectral regions, each 10 
to 20 \AA\ wide, centered at $\lambda \simeq 1290$\,\AA , $1340$\,\AA , 
$1450$\,\AA , $1700$\,\AA , $1830$\,\AA , and $1960$\,\AA\ which are free of 
detectable emission lines.
The spectral indices $\alpha $ are given for each quasar in Table 2.
The properly scaled Fe emission was subtracted using an emission template
which accounts for both \feii\ and \feiii\ emission (Vestergaard \& Wilkes 
2001). 
This improves especially the measurement of \niii ]$\lambda 1750$ and reduces 
the flux of the $\lambda 1600$\,\AA\ feature (Laor et al.\,1994). 

   \begin{figure}
   \centering
   \includegraphics[width=8cm]{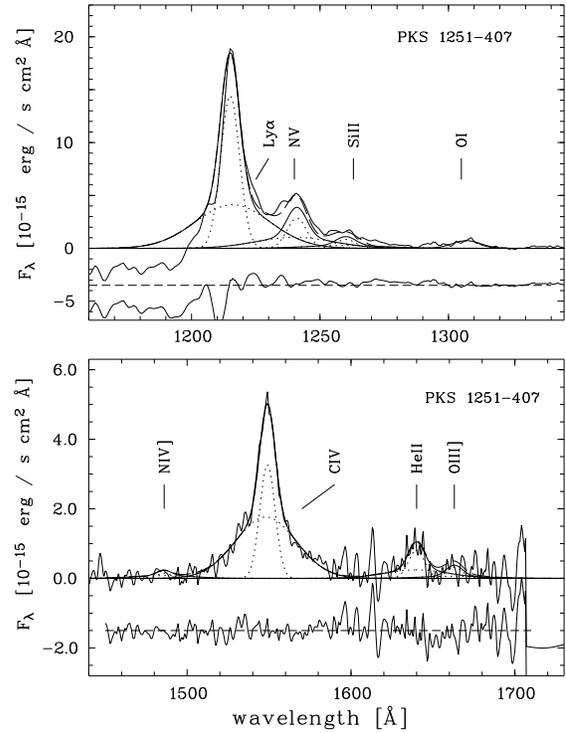}
      \caption{In the top panel an example of the Ly$\alpha $, 
         \nv $\lambda 1240$ emission line profile complex reconstruction is
         shown for PKS\,1251-407. The individual 
         profiles are displayed as solid lines while the broad and narrow 
         components are shown as dotted lines. The sum of the profile fits of 
         the individual lines is plotted as short dashed line. At the bottom 
         of the figure we plotted the difference between the individual 
         profile fits and the power-law continuum corrected quasar spectrum.
         In the bottom panel, the reconstruction of the \civ $\lambda 1549$, 
         \heii $\lambda 1640$, \oiii ]$\lambda 1663$ emission line profile 
         complex is displayed for PKS\,1251-407.}
         \label{Fig2}
   \end{figure}

We used the \civ $\lambda 1549$ emission line profile as a template to measure 
the other emission line fluxes. The \civ $\lambda 1549$ line profile was 
fitted with a 
broad and a narrow Gaussian component. We fixed the spectral width of the 
broad and narrow component in velocity space and allowed the strengths to vary 
independently. Furthermore, shifts in velocity space of the broad and narrow 
component were restricted to a range of less than a few 100\,km\,s$^{-1}$ with
respect to \civ $\lambda 1549$.
Using this \civ $\lambda 1549$ emission template is particularly important for
the \nv $\lambda 1240$ and \heii $\lambda 1640$ profiles which are blended 
with other emission lines.
To measure \ciii ]$\lambda 1909$ we used the \civ $\lambda 1549$ template to
fit \ciii ]$\lambda 1909$, \siiii ]$\lambda 1892$, and \aliii $\lambda 1857$ 
simultaneously.
This template-fitting approach can be justified since \civ $\lambda 1549$, 
\nv $\lambda 1240$, and \heii $\lambda 1640$ are all high ionization lines 
(HIL). 
Figure 2 shows typical examples of the deblending of the 
Ly$\alpha \lambda 1216$ - \nv $\lambda 1240$ and \niv ]$\lambda 1486$ -
\civ $\lambda 1549$ - \heii $\lambda 1640$ - \oiii ]$\lambda 1663$ emission 
line complexes. The resulting line flux measurements for the quasars are 
given in Table 2.
The uncertainties were estimated from the multicomponent line fit using the 
scaled \civ $\lambda 1549$ line profile to obtain a minimum $\chi^2$ fit. 
The errors are of the order of $\sim 10$\,\%\ for stronger lines like 
Ly$\alpha \lambda 1216$, \nv $\lambda 1240$, \siiv $\lambda 1400$, 
\civ $\lambda 1549$, and \ciii ]$\lambda 1909$, and $\sim 20$\,\%\ or more for
the weaker lines.
The measurement of the \niv ]$\lambda 1486$ emission line flux is severely
affected by the blue wing of the broad \civ -component. Particularly, for 
quasars with broad emission line profiles the \niv ]$\lambda 1486$ line tends 
to show a low contrast to the outer part of the \civ $\lambda 1549$ line 
profile which can be represented by the broad \civ -component only
(see Dietrich \& Hamann 2002 in prep., for further discussion).

   \begin{table*}
      \caption[]{Broad emission-line flux measurements for the observed 
              high redshift quasars. The integrated restframe emission
              line fluxes are given in units of 
              $10^{-15}$\,erg\,s$^{-1}$\,cm$^{-2}$ and $L_\lambda (1450)$ in
              erg\,s$^{-1}$\,\AA$^{-1}$.
              In addition, the spectral index $\alpha $ of the power-law
              continuum fit is given ($F_\nu \sim \nu^\alpha$).}
         \label{fluxes}
     $$ 
   \begin{tabular}{lccccccc}
            \hline
            \noalign{\smallskip}
&\multicolumn{6}{c}{$F_{rest}$[$10^{-15}$\,erg\,s$^{-1}$\,cm$^{-2}$]}\\
            \noalign{\smallskip}
            \hline
            \noalign{\smallskip}
 &Q\,0046-293&Q\,0101-304&SDSS\,0338+0021&PKS\,1251-407&APM\,1335-0417&BRI\,1500+0824\\
            \noalign{\smallskip}
            \hline
            \noalign{\smallskip}
log\,L$_\lambda (1450)$&43.60&43.59&43.45&43.32&43.56&43.49\\
            \noalign{\smallskip}
            \hline
            \noalign{\smallskip}
\ovi $\lambda 1034$   &...&$289.0\pm26.2$&...&$82.7\pm9.8$&...&$62.3\pm15.6$\\
Ly$\alpha 1216$       &$505\pm88$&$850\pm264$&$290\pm75$&$282.6\pm42.4$&$480\pm96$&$360\pm46.1$\\
\nv $\lambda 1240$    &$ 90\pm18$    &$210\pm53$    &$83.1\pm20.8$&$65.6\pm8.8$  &$85\pm17$     &$50.0\pm7.5$\\
\Sizw $\lambda $1263  &$ 25\pm 2$    &$35\pm9$      &$6.0\pm1.5$  &$18\pm3.0$    &$12.0\pm2.5$  &$13.0\pm2.0$\\
\oi $\lambda 1305$    &$ 22.4\pm0.6$ &$28.4\pm7.1$  &$13.0\pm2.5$ &$9.9\pm1.0$   &$10.5\pm1.5$  &$11.0\pm1.8$\\
\cii $\lambda 1335$   &$ 15.2\pm0.7$ &$16.2\pm4.1$  &$15.0\pm3.1$ &$2.0\pm0.9$   &$8\pm1.4$     &$8.0\pm1.7$\\
\siiv $\lambda 1402$  &$ 79.4\pm8.5$ &$109.8\pm22.0$&$60.0\pm8$   &$25.7\pm1.3$  &$53.5\pm9.8$  &$45.3\pm5.8$\\
\niv ]$\lambda 1486$  &...&$13.7\pm1.8$  &...      &$5.3\pm1.6$   &$12.0\pm2.5$  &$6.3\pm1.4$  \\
\civ $\lambda 1549$   &$163.3\pm28.6$&$276.1\pm85.7$&$170\pm43$   &$120.0\pm9.4$ &$143.5\pm29.6$&$101.8\pm13.0$\\
\heii $\lambda 1640$  &$ 25.5\pm2.2$ &$45\pm9$      &$40\pm8$     &$22.0\pm3.0$  &$27.9\pm5.6$  &$17.0\pm3.4$\\
\oiii ]$\lambda 1663$ &$ 17.9\pm1.9$ &$33\pm7$      &...   &$7.0\pm2.0$   &$17.5\pm3.5$&$9.5\pm2.6$\\
\niii ]$\lambda 1750$ &$ 18.0\pm2.7$ &$40\pm8$      &...   &...    &...    &$12.5\pm2.5$\\
\aliii $\lambda 1857$ &$ 27.0\pm4.1$ &...    &...   &...   &...    &$10.0\pm2.0$\\
\siiii ]$\lambda 1892$&$ 10.0\pm2.0$ &$20\pm5$      &...   &...    &...    &$21.8\pm2.9$\\
\ciii ]$\lambda 1909$ &$ 65.0\pm9.9$ &$170\pm35$    &...   &...    &...    &$65.5\pm12.8$\\
            \noalign{\smallskip}
            \hline
            \noalign{\smallskip}
$\alpha $&$0.09\pm0.06$&$-0.24\pm0.14$&$-0.01\pm0.20$&$-0.11\pm0.17$&$-0.45\pm0.07$&$-0.57\pm0.06$\\
            \noalign{\smallskip}
            \hline
            \noalign{\smallskip}
  &BRI\,1557+0313&Q\,2133-4311&Q\,2133-4625&Q\,2134-4521&PC\,2331+0216& \\
            \noalign{\smallskip}
            \hline
            \noalign{\smallskip}
log\,L$_\lambda (1450)$&43.21&42.82&42.78&43.15&43.43& \\
            \noalign{\smallskip}
            \hline
            \noalign{\smallskip}
\ovi $\lambda 1034$  &$83.5\pm14.2$&$34.8\pm6.3$&$7.38\pm1.40$&$58.1\pm12.2$&$130.6\pm25.7$& \\
Ly$\alpha 1216$      &$275\pm20.6$ &$166\pm25$  &$18.9\pm5.6$ &$225\pm45$   &$415\pm62$   & \\
\nv $\lambda 1240$   &$45.5\pm6.8$ &$27.7\pm5.5$&$3.4\pm0.5$&$34.2\pm6.8$ &$89\pm13.4$  & \\
\Sizw $\lambda 1263$ &$2.0\pm0.6$  &$4.7\pm0.9$ &$1.3\pm0.3$  &$7.0\pm1.4$  &$37\pm6.5$   & \\
\oi $\lambda 1305$   &$17.5\pm2.6$ &$2.5\pm0.4$ &...  &$9.0\pm1.8$  &$20.5\pm3.1$ & \\
\cii $\lambda 1335$  &...   &...  &...   &...   &$7.5\pm1.2$  & \\
\siiv $\lambda 1402$ &$10.3\pm1.4$ &$5.5\pm0.8$ &$2.3\pm0.6$&$35.4\pm4.8$ &$42.0\pm6.5$ & \\
\niv ]$\lambda 1486$ &$5.7\pm1.2$  &$3.6\pm1.0$ &...      &$8.1\pm2.1$  &$3.4\pm1.2$& \\
\civ $\lambda 1549$  &$128.8\pm9.7$&$62.9\pm9.5$&$8.14\pm2.49$&$85.9\pm21.5$&$88.2\pm9.2$ & \\
\heii $\lambda 1640$ &$30.0\pm4.5$ &$13.4\pm4.0$&...   &$11.4\pm3.4$ &$13.5\pm2.0$ & \\
\oiii ]$\lambda 1663$&$14.0\pm2.1$ &$5.8\pm1.8$ &...   &$8.0\pm2.4$  &$9.0\pm1.4$  & \\
\niii ]$\lambda 1750$&$14.5\pm2.2$ &...  &...   &...   &$8.3\pm1.0$ & \\
\aliii $\lambda 1857$&$8.0\pm1.5$  &...  &...   &...   &...   & \\
\siiii ]$\lambda 1892$&$10.0\pm1.7$&...  &...   &...   &$14.3\pm2.2$ & \\
\ciii ]$\lambda 1909$&$69.5\pm10.5$&...  &...   &...   &$42.7\pm6.4$ & \\
            \noalign{\smallskip}
            \hline
            \noalign{\smallskip}
$\alpha $  &$-0.12\pm0.05$&$-0.11\pm0.20$&$0.71\pm0.47$&$-0.15\pm0.23$&$-0.74\pm0.04$& \\
            \noalign{\smallskip}
            \hline
\end{tabular}
     $$ 
   \end{table*}

The \ovi $\lambda 1034$ emission line flux given in Table 2 has been corrected
for Ly$\alpha$ forest absorption.
For this correction we assumed an intrinsic continuum slope of 
$\alpha = -1.76$ for $\lambda \leq 1200$\,\AA\ (Telfer et al.\,2002).
The correction factor follows by simply assuming the same fraction of the 
continuum and \ovi $\lambda 1034$ emission line flux were absorbed by the 
Ly$\alpha $ forest.
The correction factors ranged from $1.33$ to $2.67$, with an average of 
$1.71 \pm 0.42$.

Hamann et al.\,(2002) presented new results of the dependence of emission line
ratios from metallicities and the shape of the ionizing input continuum.
They calculated detailed photoionization models for a wide range of density, 
$n_e$, and continuum strength, $\phi _H$, using CLOUDY (Ferland et al.\,1998). 
To study the influence of the chemical composition and of the continuum shape, 
the metallicity was varied from $Z/Z_\odot = 0.2$ to $10$ with three different 
input continua --- a broken power-law continuum with a UV bump 
(Mathews \& Ferland 1987), a single power-law continuum with $\alpha = -1$,
and a segmented power-law based on 
Zheng et al.\,(1997) and Laor et al.\,(1997). 
For more details of the model calculations see Hamann et al.\,(2002).
We used the results of these calculations to derive the chemical composition 
of the line emitting gas. 
The most reliable line ratios are those of \niii ]/\oiii ] and 
\nv/(\ovi $+$\civ).
In general, the derived metallicities depend only weakly on 


   \begin{figure*}
   \centering
   \includegraphics[width=145mm]{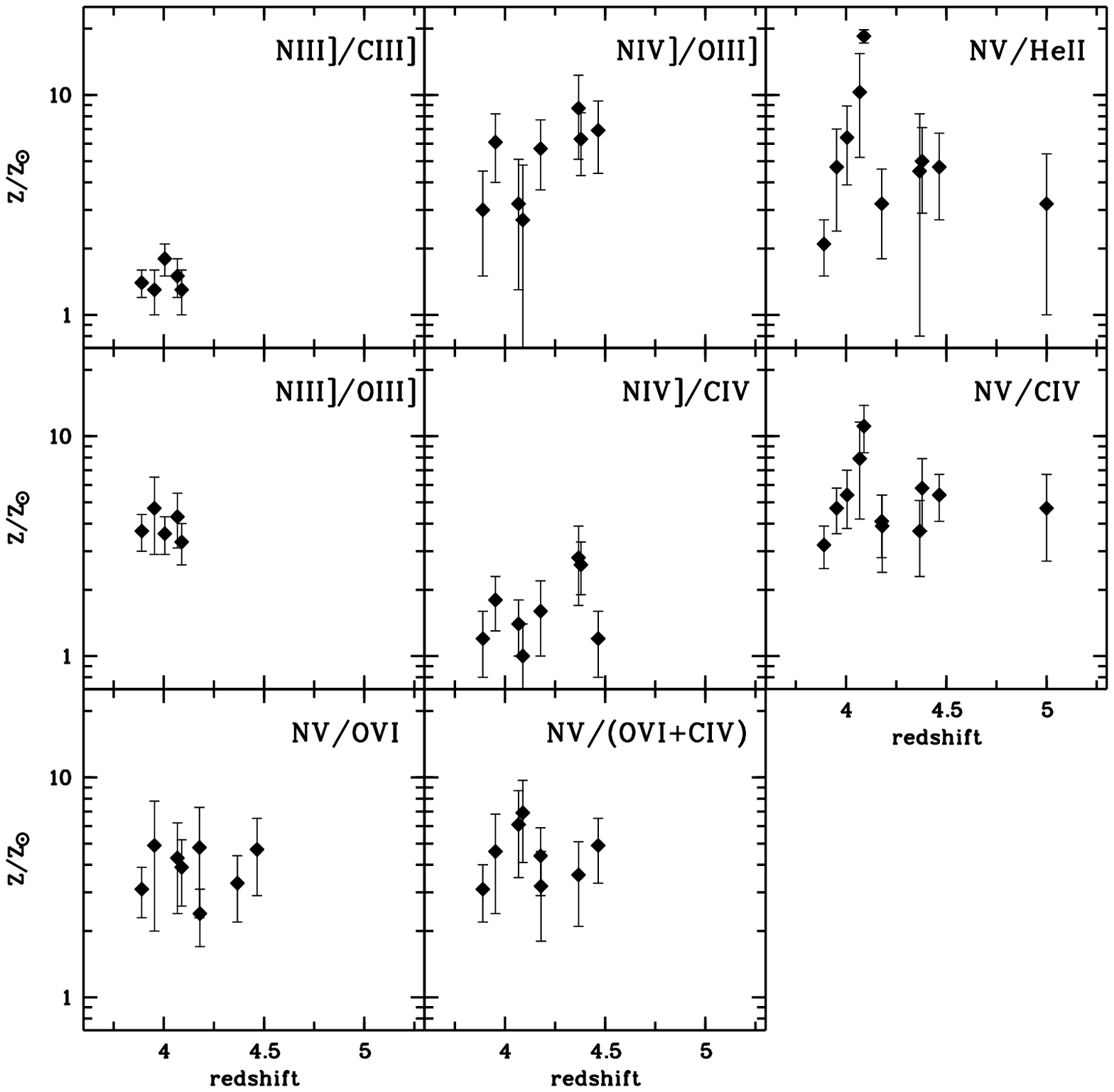}
   \caption{The metallicities of the individual high redshift quasars as a 
         function of redshift. The chemical composition which is derived 
         from the emission line ratios is plotted for each ratio.}
            \label{Fig3}
    \end{figure*}

\noindent
the shape of the ionizing continuum. 
The strongest influence of the input continuum shape is found for \nv /\heii,
where we find a factor of 2 difference depending on the continuum.
Our estimates of the metallicity for the high redshift quasars under study 
(Table 3) are based on model results using the segmented power-law continuum,
mentioned above.
The metallicities based on the other spectral shapes deviate $\la 25$\,\%\ 
from the values derived for a broken power-law continuum.
The largest deviations occur for the \nv /\heii\ line ratio.
This line ratio is quite sensitive to the temperature of the gas and on the 
ionizing continuum shape. However, Hamann \& Ferland (1993) and
Ferland et al.\,(1996) showed that this line ratio provides a firm lower 
limit on N/He when adopting a BELR ionization parameter that maximizes the 
predicted \nv /\heii\ line ratio, e.g. a hard power-law continuum 
($\alpha = -1.0$).

Ratios of the line fluxes in Table 2 were transformed to metallicity 
estimates of the gas using Figure 5 in Hamann et al.\,(2002).
The chemical composition of the gas as provided by each of the emission line 
ratios is plotted as a function of redshift in Figure 3. 
The most suited line ratios, \niii ]/\oiii ] and \nv /(\ovi $+$\civ ), as well
as \nv /\ovi, \nv /\civ , yield consistent metallicity estimates. 
The errors of the metallicities, given in Table 3, were derived from the
errors of the line flux measurements. These uncertainties were used to
calculate the uncertainty of the line ratios which yields the range of 
metallicities for a given line ratio, using the results presented in 
Hamann et al.\,(2002).

In particular, gas metallicities based on \niii ]/\oiii ] and 
\nv /(\ovi$+$\civ) agree within $\sim 33$\,\%.
The ratio \niii ]/\ciii ] tends to indicate lower metallicities than the other
line ratios. This may be due to the fact that the ratio depends more on the 
temperature of the gas than \niii ]/\oiii ]. Furthermore, the critical 
densities of \niii ] and \ciii ] differ by more than a factor of 2. Hence,
this line ratio is not as robust as \niii ]/\oiii ], since uncertainties are
introduced if the emission is received from spatially different parts of the 
line emitting region with different physical conditions (Hamann et al.\,2002).
Overall we conclude that the line ratios involving \niii ] and \nv\ provide 
quite consistent estimates of the gas metallicity for quasars.

   \begin{table*}
      \caption[]{Estimated relative abundance of the line emitting gas given 
              in units of solar metallicity Z$_\odot$.
              The estimates based on the results using a segmented power-law
              continuum fit for the photoionization models 
              (Hamann et al.\,2002).}
         \label{metallicity}
     $$ 
   \begin{tabular}{lccccccccc}
            \hline
            \noalign{\smallskip}
quasar&\multicolumn{6}{c}{$Z/Z_\odot$}\\
            \noalign{\smallskip}
            \hline
            \noalign{\smallskip}
&\hspace*{-1.5mm}\niii ]/\oiii ]\hspace*{-1.5mm}&\hspace*{-1.5mm}\niii ]/\ciii]\hspace*{-1.5mm}&\hspace*{-1.5mm}\niv ]/\civ\hspace*{-1.5mm}&\hspace*{-1.5mm}\niv ]/\oiii ]\hspace*{-1.5mm}&\hspace*{-1.5mm}\nv /\heii\hspace*{-1.5mm}&\hspace*{-1.5mm}\nv /\civ\hspace*{-1.5mm}&\hspace*{-1.5mm}\nv /\ovi\hspace*{-1.5mm}&\hspace*{-1.5mm}\nv /(\ovi$+$\civ)\hspace*{-1.5mm}&mean\\
            \noalign{\smallskip}
            \hline
            \noalign{\smallskip}
Q 0046-293&$3.6^{+0.7}_{-0.6}$&$1.8^{+0.3}_{-0.3}$&...&...&$ 6.4^{+2.5}_{-2.0}$&$ 5.4^{+1.6}_{-1.5}$&...&...&$4.3\pm0.7$\\
Q 0101-304&$4.3^{+1.2}_{-1.2}$&$1.5^{+0.3}_{-0.3}$&$1.4^{+0.4}_{-0.4}$&$3.2^{+1.9}_{-1.6}$ &$10.3^{+5.1}_{-5.1}$&$7.9^{+3.7}_{-3.4}$&$4.3^{+1.9}_{-1.3}$&$6.1^{+2.6}_{-2.2}$&$4.9\pm0.9$\\
SDSS 0338+0021&...&...&...&...&$3.2^{+2.2}_{-1.9}$&$4.7^{+2.0}_{-1.9}$&...&...&$4.0\pm1.4$\\
PKS 1251-407  &...&...&$1.2^{+0.4}_{-0.4}$&$6.9^{+2.5}_{-3.1}$&$4.7^{+2.0}_{-0.7}$&$5.4^{+1.3}_{-1.0}$&$4.7^{+1.8}_{-1.0}$&$4.9^{+1.6}_{-1.1}$&$4.6\pm0.6$\\
APM 1335-0417 &...&...&$2.6^{+0.7}_{-0.8}$&$6.3^{+2.0}_{-2.7}$& $5.0^{+2.1}_{-1.5}$&$5.8^{+2.1}_{-1.8}$&...&...&$4.9\pm0.9$\\
BRI 1500+0824 &$4.7^{+1.8}_{-1.6}$&$1.3^{+0.3}_{-0.3}$&$1.8^{+0.5}_{-0.5}$&$6.1^{+2.1}_{-2.9}$&$4.7^{+2.3}_{-1.3}$&$4.7^{+1.1}_{-1.1}$&$4.9^{+2.9}_{-1.6}$&$4.6^{+2.2}_{-1.6}$&$4.1\pm0.6$\\
BRI 1557+0313 &$3.7^{+0.7}_{-0.7}$&$1.4^{+0.2}_{-0.2}$&$1.2^{+0.4}_{-0.4}$&$3.0^{+1.5}_{-1.4}$&$ 2.1^{+0.6}_{-0.4}$&$ 3.2^{+0.7}_{-0.6}$&$3.1^{+0.8}_{-0.8}$&$3.1^{+0.9}_{-0.8}$&$2.6\pm0.3$\\
Q 2133-4311   &...&...&$1.6^{+0.6}_{-0.6}$&$5.7^{+2.0}_{-3.2}$&$ 3.2^{+1.4}_{-1.3}$&$ 4.1^{+1.3}_{-1.1}$&$4.8^{+2.5}_{-1.5}$&$4.4^{+1.5}_{-1.4}$&$4.0\pm0.7$\\
Q 2133-4625   &...&...&...&...&...&$ 3.9^{+1.5}_{-1.4}$&$2.4^{+0.7}_{-0.7}$&$3.2^{+1.4}_{-1.4}$ &$3.2\pm0.7$\\
Q 2134-4521   &...&...&$2.8^{+1.1}_{-1.1}$&$8.7^{+3.6}_{-2.9}$&$ 4.5^{+3.7}_{-1.9}$&$ 3.7^{+1.4}_{-1.3}$&$3.3^{+1.1}_{-1.1}$&$3.6^{+1.5}_{-1.5}$&$4.4\pm0.8$\\
PC 2331+0216  &$3.3^{+0.7}_{-0.6}$&$1.3^{+0.3}_{-0.3}$&$1.0^{+0.4}_{-0.3}$&$2.7^{+2.1}_{-1.8}$&$18.5^{+1.3}_{-2.0}$&$11.1^{+2.7}_{-2.7}$&$3.9^{+1.3}_{-1.1}$&$6.9^{+2.8}_{-2.3}$&$6.1\pm0.6$\\
            \noalign{\smallskip}
            \hline
            \noalign{\smallskip}
mean &$3.9\pm0.5$&$1.5\pm0.1$&$1.7\pm0.2$&$5.3\pm0.9$&$6.3\pm0.8$&$5.5\pm0.6$&$3.9\pm0.5$&$4.6\pm0.6$& \\
            \noalign{\smallskip}
            \hline
\end{tabular}
     $$ 
   \end{table*}

For a few quasars we could also compare the metallicities based on 
\niv ]/\oiii ] with the results using ratios including \niii ] and \nv ,
respectively. The metallicities we obtained using \niv ]/\oiii ] are in
agreement with the chemical composition of the gas based on
\niii] and  \nv\ line ratios to within $\sim 50$\,\%.
However, the metallicities obtained by analyzing \niv ]/\civ\ are lower than 
the other ratios (Table 3, Figure 3).
The tendency to significantly lower metallicity estimates based on 
\niv ]/\civ\ was also noted by Shemmer \& Netzer (2002). 
This is presumably due mostly to the difficulty in measuring the line flux of 
the weak \niv ]$\lambda 1486$ emission line. 
In particular, for quasars with broad emission line profiles the 
\niv ]$\lambda 1486$ line is located in the outer wing of the 
\civ $\lambda 1549$ profile. With a typical strength of about $\sim 5$\,\%\ of
the \civ $\lambda 1549$ line flux for solar metallicities, this line can be 
well hidden in the outer 
blue wing of \civ . 
The physical reason for the discrepancy is not understood.
However, it is important to keep in mind that this emission line ratio 
compares an intercombination line to a strong permitted line which may 
originate under different physical conditions.

For each quasar, we used all of the available metallicity estimates based on 
the individual emission line ratios to calculate the mean metallicity. 
With the  exceptions of SDSS\,0338+0021 and Q\,2133-4625, 4 to 8 individual 
estimates were averaged.
The resulting metallicities are typically several times solar (Table 3).
The mean metallicity for each quasar, given as the average of the 
individual estimates, is shown as a function of redshift in Figure 4.
The overall average gas chemical composition for the 11 high redshift quasars,
using these mean metallicity estimates, is $Z/Z_\odot = 4.3\pm0.3$.
\footnote{Recent studies on the photospheric solar abundance indicate 
          carbon and oxygen are about 30\,\%\ lower 
          than the values given by Grevesse \& Sauval (1998), while the 
          nitrogen abundance remains nearly unchanged within the errors
          (Holweger 2001; Allende Prieto et al.\,2001).
          Hence, this might result in a reduced estimate of the supersolar 
          abundances for the high redshift quasars by $\sim 30$\,\%.}
We also calculated the mean metallicity given by each line ratio
(Table 3, last line). The most robust metallicity indicators, \niii ]/\oiii ]
and \nv/(\ovi$+$\civ ), infer quite consistent metallicities for this small
sample of high-z quasars. Although the less suited line ratios, \niii ]/\ciii]
and \niv ]/\civ , indicate a lower metallicity, it is still at least solar.

The emission line ratios indicate that the gas chemical composition of the 
BELR at redshifts $z\ga 4$ is several times solar and at least of solar 
metallicity.
In an ongoing study we are investigating the reason for the lower metallicities
inferred using \niv ]/\civ\ and \niii] /\ciii ] compared to the robust line 
ratios \niii ]/\oiii ] and \nv/(\ovi$+$\civ ) to obtain solid measurements
of the metallicity.

\section{Discussion}

The supersolar metallicities we derived for the emission line gas in quasars 
at $z \ga 4$ provide valuable information about the preceding star formation 
epoch.
This early star formation epoch may well correspond to the beginning of major
star formation in the host galaxies. In the context of one-zone chemical 
evolution models (e.g., Hamann \& Ferland 1992,\,1993; 
Padovani \& Matteucci 1993; Matteucci \& Padovani 1993), as well as 
in more recent multi-zone models (Gnedin \& Ostriker 1997;
Fria\c{c}a \& Terlevich 1998; Granato et al.\,2001; Romano et al.\,2002)
which take into account the feedback of the dynamical and chemical evolution 
of the forming galaxies, super-solar metallicities of the gas closely related 
to quasars at high redshifts ($z\ga 4$) can be expected. 
To achieve the observed high metallicities, the single-zone and multi-zone 
models indicate evolutionary time scales for a major star formation episode of 
$\tau _{evol} \simeq 0.5 - 0.8$\,Gyrs.

   \begin{figure}
   \centering
   \includegraphics[width=8cm]{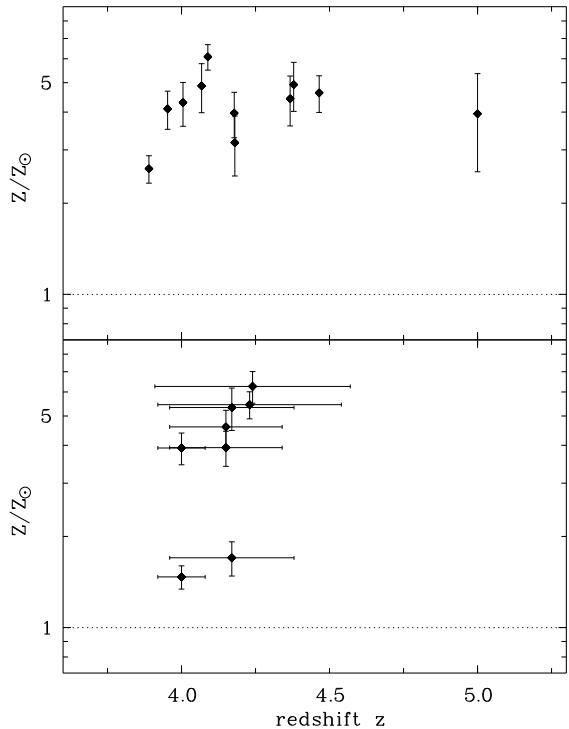}
      \caption{In the top panel, the average metallicity of the individual 
         high redshift quasars is shown as a function of redshift. The dashed 
         line marks solar metallicity $Z_\odot$. In the bottom panel the mean
         metallicity for each individual line ratio is displayed together with
         the mean redshift range covered by the quasars for which the
         individual line ratios can be measured.}
         \label{Fig4}
   \end{figure}

Based on the overall mean metallicities of $Z/Z_\odot = 4.3 \pm 0.3$ of the 
high-redshift quasars in this study ($z \ga 4$) and the time scale for the star
formation necessary to enrich the gas, the epoch of the first star formation 
can be estimated. A redshift $z\ga 4$ corresponds to an age of the 
universe of less than $\sim 1.3$\,Gyrs 
($H_o = 65$\,km\,s$^{-1}$\,Mpc$^{-1}$, $\Omega _M = 0.3$,
$\Omega _\Lambda = 0.7$). An evolution time scale of
$\tau _{evol} \simeq 0.5 - 0.8$\,Gyrs implies that the first major star 
formation in the quasars studied here must have started at 
$z_f \simeq 6 - 8$.
It is interesting to note that this is also the epoch that is supposed to mark
the re-ionization of the universe
(Haiman \& Loeb 1998; Becker et al.\,2001; Fan et al.\,2002).

Comparable high metallicities were measured for a small sample of quasars at 
$z \simeq 2.4 - 3.8$ (Dietrich et al.\,1999; Dietrich \& Wilhelm-Erkens 2000), 
for a large sample with $z \simeq 2 - 3$ (Hamann \& Ferland 1993), and for
70 quasars with $z\ga 3.5$ (Dietrich et al.\,2002, in prep.). In 
particular, there is no decline in metallicity from $z\simeq 2$ to $z\ga 4$.
This can be taken as an indication that the formation of massive spheroidal
systems, accompanied by intense star formation, starts at $z_f \simeq 6 - 8$ 
and continues until $z\simeq 2$ (Madau et al.\,1996; Steidel et al.\,1999).
At the end of the first major star formation episode, which lasts for 
$\sim 0.5 - 0.8$\,Gyrs, quasar activity starts in environments that are already
highly enriched (e.g., Granato et al.\,2001; Romano et al.\,2002). 

\section{Conclusion}

We observed a sample of 11 high redshift quasars with $3.9 \la z \la 5$ at low
spectral resolution.
We used several emission-line fluxes ratios involving carbon, nitrogen, 
oxygen, and helium to estimate the metallicity of the line emitting gas. 
To transform the observed line ratios into metallicities we used the
results of detailed photoionization calculations (Hamann et al.\,2002).
The emission line ratios involving \niii ] and \nv\ provide generally 
consistent estimates of the gas metallicity for quasars. 
In particular, the results based on \niii ]/\oiii ], \nv /(\ovi $+$\civ), 
\nv /\ovi , and \nv /\civ\ differ by less than $\sim 30$\,\%.
The average metallicity for the 11 high redshift quasars in our sample is 
$Z/Z_\odot = 4.3 \pm 0.3$.
We placed these results in the context of chemical evolution models 
presented by Hamann \& Ferland (1993) and Fria\c{c}a \& Terlevich (1998).
For an evolution/enrichment time scale of approximately 
$\tau _{evol} \sim 0.5 ~{\rm to}~ 0.8$\,Gyrs, we estimate that the first 
major star formation must have begun in these environments at a redshift of 
$z_f \simeq 6 ~{\rm to}~ 8$, i.e., at a cosmic age of less than 1\,Gyr 
($H_o = 65$ km\,s$^{-1}$\,Mpc$^{-1}$, $\Omega _M = 0.3$, 
 $\Omega _\Lambda = 0.7$).

\begin{acknowledgements}
      We are grateful to our colleagues J.A.\,Baldwin, G.J.\,Ferland, and 
      K.T.\,Korista for helpful discussions.
      MD and FH acknowledge support from NASA grant NAG 5-3234 and  NSF grant 
      AST-99-84040 (University of Florida). JH and MD were also supported by 
      the grants SFB~328~D and SFB\,439 (Landessternwarte Heidelberg).
      MV gratefully acknowledges financial support from the Columbus 
      Fellowship.
\end{acknowledgements}



\begin{thebibliography}{}

\bibitem[]{}
        Allende Prieto, C., Lambert, D.L., \& Asplund, M. 2001, \apj, 556, L63
\bibitem[]{}
        Andreani, P., La Franca, F., \& Cristiani, S. 1993, \mnras, 261, L35
\bibitem[]{}
        Arimoto, N. \& Yoshii, Y. 1987, \aap, 173, 23
\bibitem[]{}
        Baldwin, J.A. \& Netzer, H. 1978, \apj, 226, 1
\bibitem[]{}
        Becker, R.H., et al. 2001, \aj, 122, 2850
\bibitem[]{}
        Burstein, D. \& Heiles, C. 1982, \aj, 87, 1165
\bibitem[]{}
        Carilli, C.L., Bertoldi, F., Omont, A., Cox, P., McMahon, R.G., \&
        Isaak, K.G. 2001, \aj, 122, 1679
\bibitem[]{}
        Cen, R. \& Ostriker, J.P. 1999, \apj, 519, L109
\bibitem[]{}
        Connolly, A.J., Szalay, A.S., Dickinson, M., Subbarao, M., \& Brunner,
        R.J. 1997, \apj, 486, L11
\bibitem[]{}
        Davidson, K. 1977, \apj, 218, 20
\bibitem[]{}
        Dickinson, M. 1998, in ``The Hubble Deep Field'', STScI Symp., eds.
        M.\,Livio, S.\,Fall, \& P.\,Madau, p.219
\bibitem[]{}
        Dietrich, M., Appenzeller, I., Vestergaard, M., \& Wagner, S.J. 2002a, 
        \apj, 564, 581
\bibitem[]{}
        Dietrich, M., Appenzeller, I., Wagner, S.J., et al. 1999, \aap, 352, L1
\bibitem[]{}
        Dietrich, M., et al. 2002b, \apj, 581, in press
\bibitem[]{}
        Dietrich, M. \& Wilhelm-Erkens, U. 2000, \aap, 354, 17
\bibitem[]{}
        Fan, X., et al. 1999, \aj, 118, 1
\bibitem[]{}
        Fan, X., Narayanan, V.K., Strauss, M.A., White, R.L., Becker, R.H., 
        Pentericci, L., \& Rix, H.-W. 2002, \aj, 123, 1247
\bibitem[]{}
        Ferland, G.J., Baldwin, J.A., Korista, K.T., Hamann, F., 
        Carswell, R.F., Phillips, M., Wilkes, B., \& Williams, R.E. 1996, 
        \apj, 461, 683
\bibitem[]{}
        Ferland, G.J., Korista, K.T., Verner, D.A., Ferguson, J.W., Kingdon,
        J.B., \& Verner, E.M. 1998, \pasp, 110, 761
\bibitem[]{}
        Fria\c{c}a, A.C.S. \& Terlevich, R.J. 1998, \mnras, 298, 399
\bibitem[]{}
        Gaskell, C.M., Shields, G.A., \& Wampler, E.J. 1981, \apj, 249, 443
\bibitem[]{}
        Gebhardt, K., et al. 2000, \apj, 543, L5
\bibitem[]{}
        Gnedin, N.Y. \& Ostriker, J.P. 1997, \apj, 486, 581
\bibitem[]{}
        Granato, G.L., Silva, L., Monaco, P., Panuzzo, P., Salucci, P., De 
        Zotti, G., \& Danese, L. 2001, \mnras, 324, 757
\bibitem[]{}
        Grevesse, N. \& Sauval, A.J. 1998, Space Sci.Rev., 85, 161
\bibitem[]{}
        Haiman, Z. \& Loeb, A. 1998, \apj, 503, 505
\bibitem[]{} 
        Hamann, F. 1997, \apjs, 109, 279
\bibitem[]{}
        Hamann, F. \& Ferland, G.J. 1992, \apj, 381, L53
\bibitem[]{}
        Hamann, F. \& Ferland, G.J. 1993, \apj, 418, 11
\bibitem[]{}
        Hamann, F. \& Ferland, G.J. 1999, \araa, 37, 487
\bibitem[]{}
        Hamann, F., Korista, K.T., Ferland, G.J., Warner, C., \& Baldwin, J.A.
        2002, \apj, 564, 592
\bibitem[]{}
        Hamuy, M., Walker, A.R., Suntzeff, N.B., Gigoux P., Heathcote, S.R., \&
        Phillips, M.M. 1992, \pasp, 104, 533
\bibitem[]{}
        Holweger, H. 2001, in Joint SOHO/ACE workshop ``Solar and Galactic
        Composition'', ed. R.F.\,Wimmer-Schweingruber, AIP Conf.Proc.\,Vol.598,
        p.23
\bibitem[]{}
        Horne, K. 1986, \pasp, 98, 609
\bibitem[]{}
        Isaak, K.G., McMahon, R.G., Hils, R.E., \& Withington, S. 1994, 
        \mnras, 269, L28
\bibitem[]{}
        Izotov, Y.I. \& Thuan, T.X. 1999, \apj, 511, 639
\bibitem[]{}
        Kaspi, S., Smith, P.S., Netzer, H., Maoz, D., Jannuzi, B.T., \& Giveon,
        U. 2000, \apj, 533, 631
\bibitem[]{}
        Kauffmann, G. \& Haehnelt, M.G. 2000, \mnras, 311, 576
\bibitem[]{}
        Kormendy, J. \& Richstone, D. 1995, \araa, 33, 581
\bibitem[]{}
        Kormendy, J. \& Gebhardt, K. 2001, in AIP conf.proc. Vol.586,
        20th Texas Symposium on relativistic Astrophysics XIX, eds. 
        J.C.\,Wheeler \& H.\,Martel, p.363
\bibitem[]{}
        Laor, A., et al. 1994, \apj, 420, 110
\bibitem[]{}
        Laor, A., Fiore, F., Elvis, M., Wilkes, B.J., \& McDowell, J.C. 1997, 
        \apj, 477, 93
\bibitem[]{}
        Lilly, S.J., Le F\`evre, O., Hammer, F., \& Crampton, D. 1996,    
        \apj, 460, L1
\bibitem[]{}
        Madau, P., Ferguson, H.C., Dickinson, M.E., Giavalisco, M., Steidel,
        C.C., \& Fruchter, A. 1996, \mnras, 283, 1388
\bibitem[]{}
        Magorrian, J., Tremaine, S., \& Richstone, D. 1998, \aj, 115, 2285
\bibitem[]{}
        Mathews, W.G. \& Ferland, G.J. 1987, \apj, 323, 456
\bibitem[]{}
        Matteucci, F. \& Padovani, P. 1993, \apj, 419, 485
\bibitem[]{}
        Merritt, D. \& Ferrarese, L. 2001, \apj, 547, 140
\bibitem[]{}
        M\"{o}hler, S., Seifert, W., Appenzeller, I., \& Muschielok, B. 1995, 
        in ESO Workshop ``Calibrating and Understanding HST and ESO 
        Instruments'', ed. P.\,Benvenuti, ESO, p.145
\bibitem[]{}
        Omont, A., Cox, P., Bertoldi, F., McMahon, R.G., Carilli, C., \& 
        Isaak, K.G. 2001, \aap, 374, 371
\bibitem[]{}
        Omont, A., McMahon, R.G., Cox, P., Kreysa, E., Bergeron, J., Pajot, F.,
        \& Storrie-Lombardi, L.J. 1996, \aap, 315, 1
\bibitem[]{}
        Osmer, P.S. 1980, ApJ, 237, 666
\bibitem[]{}
        Padovani, P. \& Matteucci, F. 1993, \apj, 416, 26
\bibitem[]{}
        Pagel, B.E.J. \& Edmunds, M.G. 1981, \araa , 19, 77
\bibitem[]{}
        Petitjean, P., Rauch, M., \& Carswell, R.F. 1994, \aap, 291, 29
\bibitem[]{}
        Pettini, M. 1999, in Proc.of ESO Workshop ``Chemical Evolution from 
        Zero to High Redshift'', ed.\,J.\,Walsh \& M.\,Rosa, LNP, p.233
\bibitem[]{}
        Romano, D., Silva, L., Matteucci, F., \& Danese, L. 2002, \mnras, 
        334, 444
\bibitem[]{}
        Salucci, P., Szuszkiewicz, E., Monaco, P., \& Danese, L. 1999, \mnras,
        307, 637
\bibitem[]{}
        Savage, B.D. \& Mathis, J.S. 1979, \araa, 17, 73
\bibitem[]{}
        Schwarz, H.E. \& Melnick, J. 1993, The ESO Users Manual, p.24
\bibitem[]{}
        Shemmer, O. \& Netzer, H. 2002, \apj, 567, L19
\bibitem[]{}
        Shields, G.A. 1976, \apj, 204, 330
\bibitem[]{}
        Steidel, C.C., Adelberger, K.L., Giavalisco, M., Dickinson, M., \&
        Pettini, M. 1999, \apj, 519, 1
\bibitem[]{}
        Telfer, R.C., Zheng, W., Kriss, G.A., \& Davidsen, A.F. 2002, \apj,
        565, 773
\bibitem[]{}
        Tinsley, B.M. 1979, \apj, 229, 1046
\bibitem[]{}
        Tinsley, B.M. 1980, Fundam. Cosmic Phys., 5, 287
\bibitem[]{}
        Tremaine, S., et al. 2002, \apj, 574, 740
\bibitem[]{}
        Tresse, L. \& Maddox, S.J. 1998, \apj, 495, 691
\bibitem[]{}
        Uomoto, A. 1984, \apj, 284, 497
\bibitem[]{}
        van Zee, L., Salzer, J.J., \& Haynes, M.P. 1998, \apj, 497, L1
\bibitem[]{}
        Vestergaard, M. \& Wilkes, B.J. 2001, \apjs, 134, 1
\bibitem[]{}
        Warner, C., Hamann, F., Shields, J.C., Constantin, A., Foltz, C.B., \&
        Chaffee, F.H. 2002, \apj, 567, 68
\bibitem[]{}
        Wheeler, J.C., Sneden, C., \& Truran, J.W. 1989, \araa, 27, 279
\bibitem[]{}
        Zheng, W., Kriss, G.A., Telfer, R.C., Grimes, J.P., \& Davidson, A.F.
        1997, \apj, 475, 469
\end{thebibliography}
\end{document}